\documentclass[times,5p,preprint]{myarticle}

\usepackage{amsmath,amssymb}
\usepackage{verbatim}
\usepackage{color}
\usepackage{graphicx}
\usepackage{psfrag}

\numberwithin{equation}{section}

\def\be{\begin{equation}}
\def\ee{\end{equation}}
\def\bes{\begin{equation*}}
\def\ees{\end{equation*}}
\def\sm{\smallskip\noindent}
\def\lam{\lambda}

\begin{document}

\begin{frontmatter}

\title{ Galactic exploration by directed 
self-replicating probes, \\
and its implications for the Fermi paradox} 

%\shorttitle{}

\author{
Martin T. Barlow\footnote{Research partially supported by NSERC (Canada)
and Trinity College, Cambridge}
}
\ead{barlow@math.ubc.ca}

 \address{
 Department of Mathematics,
 University of British Columbia,
 Vancouver, BC V6T 1Z2, Canada. \\
 Email: barlow@math.ubc.ca }
 
%\maketitle

\begin{abstract}
This paper proposes a long term
scheme for robotic exploration of the galaxy, and then considers
the implications in terms of the `Fermi paradox' and our search for ETI.  
We discuss the  `galactic ecology'  of civilizations in terms of the
parameters $T$ (time between ET civilizations arising) and $L$, the
lifetime of these civilizations. Six different regions are described.


\end{abstract}

\end{frontmatter}

\section{Introduction} \label{sec:intro}

The proposal to explore the galaxy by self-replicating probes goes back at
least as far as Freitas 1980a and Tipler 1980.
Numerous criticisms have been made, notably by Sagan and Newman 1983.
A recent paper (Wiley 2012) reconsiders the topic, and finds some of these
criticisms without much merit.
This paper outlines in more detail one exploration scheme and its benefits,
and then considers the implication for 
the galactic `ecology' of intelligent species.

\section { Exploration of the galaxy by self replicating probes }

Our starting point is as in  Freitas 1980b and Tipler 1980. This
relies on two technologies which we do not have
at present, but which it is reasonable to suppose we will attain within 
the next few hundred years:

\sm (T1) A propulsion system capable of sending probes 
to nearby stars, at say 0.01 c. \\
(T2) An AI system which, in total, is able, with the resources
found in  most star systems, of replicating itself.

\sm {\sl Propulsion systems.}  While no existing system can reach 
these speeds, various proposals for 100
year probes to Alpha Centauri have been made, such as Project
Daedalus. A speed of 0.01c does not seem unduly optimistic, and in
fact many proposals for galactic exploration, such Bj\o rk (2007),
consider probes which travel at 0.1c.

\sm {\sl AI systems}
Again, we do not have self-replicating systems at present, and
what such a system would be like must rely largely on conjecture. 
In this context note the following remark in Wiley (2012):
\begin{quote}
\textit{One point we take issue with is an inherent and frequently unconscious
{\it biological bias} that pervades consideration of computerized intelligence,
including SRPs.} 
\end{quote}

A common tendency has been to imagine a self-replicating machine as being
rather like a bacterium: that is a single machine which (somehow, almost
magically) is able to move around in its environment and 
replicate itself. If we start with present
technology, we are forced to imagine something rather different.
The system as a whole might consist of 3 parts:\\
(A) A number of robots and probes, of several different types, which
are together capable of exploring a solar system, and gathering 
resources (metals, volatiles etc).\\
(B) A `slow assembler' which would be able refine these materials into
components which would make the final factory (C).\\
(C) A large scale factory, or collection of factories, which would
be able to manufacture copies of (A) and (B), as well as
additional surveying and communication devices. 

The payload of the probe would consist of (A+B), together with enough 
raw materials (fuel etc.) to get started in the new system. 
Once (C) was made, resources would be gathered for as long
as was necessary, and a number of probes would then be sent to
nearby stars. If we take this view, then a `self-replicating space
probe' (SRP) would not be a single machine, but rather a 
collection of different machines with an overall capability of replication. 

See Freitas 1980b for a much more detailed description of such a probe,
with the probe (A+B) plus the fuel for the voyage having a mass of 
of around $10^{10}$ kg. The factory described
there only makes 1 new probe every 500 years, but (see Section 7.1) using
a longer period for the initial construction gives a larger factory which can
create 1000 new probes in 1500-200 years. For simplicity I will take the
reproduction time, between the arrival of the initial probe in a star system,
and the completed factory (C) sending out new probes, to be $T_R=1000$.

The AI needed for such a system far exceeds what is possible at
present.  But while the kinds of decisions necessary for the AI
(e.g. `what kind of material is present in this asteroid?', `can it be
transported to the factory?')  would require a very high level of
skill, this would be within fairly narrow parameters, and 
a human level of overall initiative and
judgement would not be required.  Even if machines with a human
intelligence can be constructed, it might be desirable to limit the
intelligence of the SRPs.

As Tipler 1980 notes, there are reasons other than
stellar exploration to develop these technologies. Progress in
AI has been far slower than supposed by early optimists, but it still seems
reasonable to suppose that, within a few hundred years, we will be able to
build such SRPs.
The development of such machines would, at least for a while, introduce an
age of plenty, since it would open up the resources of the solar system
for our exploitation. Some idealistic individuals or groups might then be
willing to invest the resources in making a number of the probes (A+B),
and send them to nearby stars. (See Mathews 2011 for another
proposal to explore the galaxy by SRPs.) 

\subsection{Exploration strategy}

I now propose a scheme for galactic exploration, assuming that we
do develop the technologies (T1) and (T2) described above. 
The first step would be to send
probes to the 10--100 nearest stars, which are all within about 20 ly of Earth.
(Long before the start of such a mission we will have very good data on the
planetary systems of these stars.)
The probes would arrive at their destinations
400-2000 years after the mission start. They would remain
in radio contact with Earth, (with a time lag of 40 years or less), 
would report on their discoveries, and would be able to receive updates 
on strategy. (Among the exploration devices (A) would be systems
able to transmit and receive narrow band radio or laser 
communication over a distance of say 100 ly.)

I will call the initial star systems Level 1 `colonies', though there is
no suggestion that they would have a human population.
After the construction of the factory (C) on a Level 1 colony, the
colony would send out SRPs (let us say about 1,000 -- 10,000) to
create colonies at `Level 2'. 
I have suggested an initial `hop size' of 10-20 ly, since the number
of probes that could be sent out from our Solar System might be limited by
resource constraints. However once a colony at Level 1 or higher had 
a working factory, there would be no such limit on the number of probes that 
could be sent out, and it would be sensible to send as many as was 
necessary to explore every star system within the second `maximum hop size'.
There are about 15,000 stars within 100 ly of Earth,
so with some useful duplication the Level 1 systems would together be able
to send probes to every star in this region.

The maximum hop size $h_m$
would be the greatest distance such a probe could be sent
with a probability greater than 90\% of arriving. I will take 
$h_m=100$; this also needs to be 
less than the maximum distance for radio or laser
communication, but this is much greater than 100 ly.

The Probes from the Level 2 colonies would then establish Level 3 colonies, 
and so on. Each colony at Level $n$ would report back to its Level $n-1$ ancestor, and
receive updating instructions from it. 
While it would be desirable for the Level 1 colonies to produce 
many probes, as the radius of
the exploration sphere became larger, and so the curvature of its surface became
less, fewer new probes per colony would be needed.

Within a few thousand years of the mission start our descendants on earth
(if they still existed) would be receiving a flood of information from
the exploration of hundreds of star systems. The Great Pyramid
was built around 4500 years ago; 4500 years after its start 
the mission would be well under way, and have given us detailed data on every
star system within about 30 ly of Earth.

The overall mission would continue until the planetary
system of every star in the galaxy had been explored.
Let $v_p =0.01c$ be the speed of the probes, and $v_e$ be the propagation speed 
of the exploration front. Then
$$ v_e = \frac{h_m}{T_R + h_m/v_p} = v_p \frac{1}{1+ v_p T_R/h_m}. $$
So $T_R =1000$ and $h_m=100$ give $v_e/v_p = 10 /11$, the exploration front
travels at nearly the same speed as the probes, and the total time to
explore the galaxy is around $10^7$ years.
This compares with exploration times of the order of $10^8$ years given
by Bj\o rk 2007, using probes which travel at 0.1c, but do not replicate.

\subsection{Refinements}

While the basic strategy is as above, it is necessary to consider a number
of refinements.

\sm {\sl (a) Resource use within system.} The best place for the
construction of the factory (C) might be the moons of a planet in the outer
reaches of the star system. Assuming a mass of $10^{10}$ kg for the probes 
(A+B) and fuel, the construction of (C) plus say 10,000
probes would use at most a handful of minor planets and comets. This
would leave plenty of material behind, even on the Level 1 colonies,
and it would not be necessary to to `strip mine' the galaxy in order
to complete this exploration.  One would only need a few
probes per star -- one plus a margin for accidents.

\sm {\sl (b) Systems with planets with life.}
In systems with planets with complex life, a different procedure should be followed. 
Two possibilities would be: \\
(i) Report, build a factory (C), explore the system thoroughly, and then 
 await instructions from Earth, \\
(ii) Report, do nothing, and await instructions from Earth.

The first and second level probes would provide enough data to
refine this strategy, at an early stage of the overall mission.  In
the very unlikely event that more than 90\% of systems have planets
with complex life, a modification of (ii) would be needed so that a
reasonable proportion of colonies did send out probes.

\sm {\sl (c) Extinction of the human race.}  Once set in motion, the
exploration could continue without any further human
intervention. However, this proposal envisions continued interaction and direction:
Earth would receive data from the probes, and based on this revised instructions
on exploration (as well as possible system upgrades) would be sent out. 
What however if humanity becomes extinct, or just loses 
interest in the mission? There are many possible procedures which
could be followed, of which the simplest are: 
(i) Continue anyway, (ii) Abandon further exploration. 
For simplicity I propose (ii), and suggest that every 100 years the 
Level 1 colonies would ask Earth  ``Shall we continue?'' 
If 1000 years went by with no positive response, the project would be mothballed,
and instructions would be sent through the
communication tree that no further SRPs were to be built.

\sm {\sl (d) Communication and direction.} 
A key part of this exploration scheme is that the SRPs are not autonomous, but 
that the whole exploration process is directed, ultimately from Earth. 
The first requirement here is that Level $n$ colonies be able to communicate
(over a distance of say 100 ly) with their Level $n-1$ ancestors. Even with
present technology, we could build transmitters and receivers 
capable of working over these distances.

While all the Level 1 colonies would send out probes, this would not be
necessary for higher level colonies, and stars could be divided into two groups.
For the first, `end nodes', a factory C would be built, the star system explored, but
no probes would be sent out. The second, `branch nodes' would send out probes.
Among the pieces of infrastructure built in each colony would be telescopes
to survey the stellar neighbourhood, and 
using this data, nearby Level $n$ colonies would coordinate the exploration
of their neighbourhoods. (Nearby colonies
would be 10-100 ly apart, so communication time would be small compared with the
time to build probes, or for the journeys.)
While the algorithm to coordinate this process may appear complicated, 
it is well within our current capabilities --
unlike the AI needed for robotic exploration of a planetary system

\sm {\sl (e) Mission creep and machine mutation.}
A widely voiced concern with SRPs has been that they might mutate, run
amok, and eat up the galaxy -- see Sagan and Newman 1983.
However Wiley 2012 argues that it should be possible to build
in sufficient reliability to avoid this outcome -- note also 
his comment above on inappropriate biological analogies. 
In the exploration scheme proposed here about 1000 generations would be needed
to explore the galaxy -- fewer if hops longer than 100 ly are feasible.
The total number of replicators (C) built would be
of the same order as the number of stars in the galaxy, that 
is about $4 \times 10^{11}$.
Wiley 2012 points out that this is much less than the number of 
cell divisions within
a human lifetime, which is of the order of $10^{16}$. 

If necessary, further steps could be taken to reduce the overall risk. In the 
initial stages of the exploration we would want every planetary system to
be explored carefully. 
However, it seems likely that after the first million or
so systems had been explored, we would have a good understanding of the
processes underlying the formation and development of planetary systems, 
and might only be interested in those systems which had life, complex life,
or other exceptional features. 
The later phases of the exploration could therefore proceed as follows.
Each `branch node' would first send out about 10 new full probes (A+B) to 
establish the next generation of colonies. Next, it would send out reduced
probes, just consisting of (A), to all the stars in its exploration patch.
These would explore the target system, and report back to the sending
branch node. Without the reproductive capacity (B)
these probes would ultimately run out of fuel and become inactive.
A full probe (A+B) would then be sent to any 
system that merited further attention. 
Assuming that `interesting' systems are rare, this modification would reduce 
the number of full replications by a factor of 1000 or so. 
Further safety
mechanisms could also be built in, such as deeply embedded software constraints
on the total number of probes that the factories (C) could make, or on the total
number of permitted generations.

\sm {\sl (f) Crossing large spaces and percolation.} 
Landis 1998 has suggested a percolation model for the spread of a species
through the galaxy, and shown that in some cases this leads to large vacant
(unexplored) regions. However bond percolation on the lattice 
is a poor model for the type of exploration proposed above, since 
each `branch node' would send out rather more than 5 probes. 
Further, the communication envisaged between colonies would
mean that colonies would become aware of interstellar voids (with no or few 
stars), and regions that, perhaps because of the failure of a number of probes,
were remaining unexplored. They could then send additional probes to explore
these regions. 
If we consider the mathematical graph whose
vertices are the stars, and join by edges all pairs of stars within 100 ly.,
then the exploration scheme proposed here will explore all stars in
the connected component containing our sun, and it seems overwhelmingly
likely that this spans most of the galaxy. 

\sm {\sl (g) System updates.}
We would want to be 
able to incorporate updates into the systems (A,B,C). It is possible that this
could be done by radio, but the available bandwidth might be too small
for the necessary amount of data. One can imagine a system of
`fast packets' -- small probes carrying data, which travel at say
0.1c between colonies with the infrastructure to send and receive them.
However one disadvantage of allowing such updates is that it would 
make the colonies more vulnerable to mutations or computer viruses.

\sm {\sl (h) Contact with extraterrestrial intelligence (ETI).} Detailed
thought would need to be given on what course of action should be taken
if either ETI were found, or traces of them. There 
would be time to refine strategies in the first few millenia of the
mission, as data on frequency and type of life in other star systems accumulated.
The number of possible actions the probes could take is large, and 
a full discussion of this is beyond the scope of this article.
The simplest (but not the quickest) option would be for the 
the probe to report back, and take no further action until instructed.

\section { Implications for SETI,  and Fermi's question}

Let us now make the hypothesis (H) that 
the technolgies (T1) and (T2) can be attained, 
and explore the consequences.
The exploration scheme outlined above, using these technologies is
one which, if it survives long enough, the human
race might adopt -- no doubt with a number of improvements.
The payoff is that with a relatively low initial cost 
our descendants would obtain detailed data about every star system in the galaxy. 
In particular they would learn how many planets support life, what kind of life
it is, and just how rare complex or intelligent life is. 

If there have been technological ETI in the galaxy, then they would
also have had this option.  So -- this is Fermi's question ``Where are
they?'' (This is often called the `Fermi paradox', but it is only a
paradox if one begins with the assumption that intelligent life is 
common. In fact we have no information on this.)

Let us recall the Drake equation, slightly modified for our purposes:
$$ N = R^{*} \cdot f_p \cdot n_e \cdot f_{\ell} \cdot f_{ci} \cdot L ; $$
here $N$ is the number of existing civilizations sending out SRPs, 
$R^*$ is the rate of star formation per year in the galaxy,
$f_p$ is the fraction of those stars that have planets,
$n_e$ is the average number of planets
that can potentially support life per star 
that has planets, $f_l$ is the fraction of these
that develop life, $f_{ci}$ is
the fraction of these that develop 
civilizations that send out SRPs, and $L$ is the average lifetime
of such civilizations. This lifetime is the time that either the
civilization itself, or its SRPs, remain active. 
(From now on, I will use the term `civilization' for `civilizations that send out SRPs').

We do have estimates of at least the order of magnitude of 
some of the early terms in this expression:
for example $R^* \simeq 7$, and data from the Kepler satellite suggests
that $f_p \simeq 0.5$, while $n_e$ is quite small. (Out of about 10,000
systems surveyed, only a handful have planets which look really promising
from the point of finding earthlike life.)
At present $f_l$ and $f_{ci}$ are utterly unknown, though estimates of $f_l$
may at some point become available via spectroscopic search for oxygen.

I have given the Drake equation in a simple form. A more realistic
equation would take account of randomness, and the fact that these factors are
not constant in time -- see for example Glade et. al. 2012. However, the
uncertainty in our knowledge of the parameters in the equation is so great
that these refinements seem to the author of this article 
to add little to what can be achieved with a simple `back of an envelope' 
calculation.

Let us now set
$$ \lam =  R^{*} \cdot f_p \cdot n_e \cdot f_{\ell} \cdot f_{ci} = N/L; $$
so that $\lam$ is the number of civilizations arising per year in the
galaxy. As an upper bound, if $f_l = f_{ci}=1$ (surely very unlikely)
and $n_e = 0.03$, we obtain $ \lam \le 0.1$. Let us set $T=1/\lam$
to be the average length of time in years between successive civilizations
arising in the galaxy; 
the estimates above suggest it is unlikely that $T < 10$. 

Let us now consider the `galactic ecology' in terms of the two
parameters $T=\lam^{-1}$ and $L$. While a better model would allow 
for randomness of $L$, a simple mean model already yields useful insights. 
Figure 1 shows a plot of $\log L$ against $\log T$.  Since the galaxy
is about $10^{10}$ years old, we have $\log L\le 10$, 
and it seems reasonable to take also $\log L \ge 2$.
The estimates above give $\log T \ge 1$. We have no
upper bound on $T$: it is not legitimate to use the Copernican principle
to assert that because there is at least one potential civilization in
the galaxy (us) then $T \le L$. Civilizations might only arise in one
galaxy in a billion, and those that arose would still observe themselves 
to be in a galaxy. In the diagram I take $1 \le \log T \le 14$.

\psfrag{logL}{$\log L$}
\psfrag{logT}{$\log T$}
\psfrag{R1}{$R_1$}
\psfrag{R2}{$R_2$}
\psfrag{R3}{$R_3$}
\psfrag{R4}{$R_4$}
\psfrag{R5}{$R_5$}
\psfrag{R6}{$R_6$}

\psfrag{S0}{$0$}
\psfrag{S5}{$5$}
\psfrag{S10}{$10$}

\begin{figure}[t]
\centering
\includegraphics[width=\linewidth]
%{fig_predShadow_t99_m0_L-1_n2_d15_Nf___2x2}
{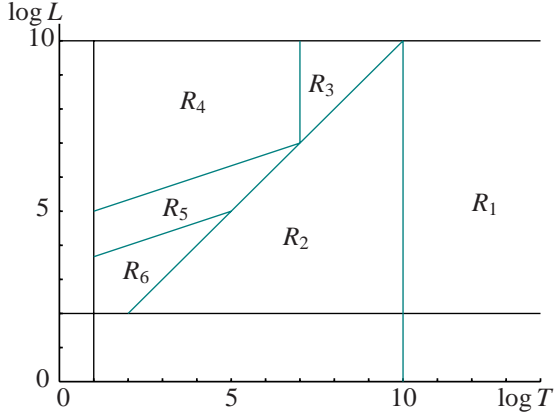}
\caption{\textbf{Galactic ecology parameter space} }
\label{G-eco}
\end{figure}

Let us now consider the various regions of the diagram.
The descriptive statements for the regions apply to typical points in the
region -- naturally these will become weaker if the point
$(\log T, \log L)$ is close to the boundary between regions.

\sm ($R_1$) (`Alone') If $\log T > 10$ then probably 
no other civilization has arisen in the galaxy. (A more accurate 
statement would be that the mean number of such civilizations is less than 1.)

\sm ($R_2$) (`Pompeii') If $\log T \le 10$ and $\log L < \log T$ then
$N < 1$ and there is no other civilization existing now. However,
$10^{10}/T \ge 1$ civilizations have existed, and their ruins
await discovery -- except that we may not last long enough to find them. 

\sm ($R_3$) (`Galactic hegemony') $\log L \ge \log T \ge 7$.
We have seen above that in a time of about $t_e=10^7$ years a civilization
can explore the galaxy via SRPs. If this civilization lasts longer than 
that, and no other civilization arises during the exploration period, then
the exploring civilization would attain `galactic hegemony'. It would know of the
existence of any other civilization that might arise, and would be able 
to control their growth and activities.

\smallskip
In the remaining parts of the diagram there are many civilizations in the galaxy.
Assume for simplicity that
the galaxy is a uniform disk of thickness $h_G=1000$ ly and radius
$R_G=50,000$ ly, that civilizations arise uniformly in the galaxy at rate
$\lam$, start exploring the galaxy by SRPs with an exploration speed of 
$v_e= 0.01c$, and continue to do so until the civilization (and the SRPs) end
$L$ years after the start of the exploration. 
(A more detailed analysis would take account of the likely
existence of a galactic habitable zone described by Lineweaver et. al. 2004.)
 
If a civilization starts at position $x_0$ and time $t_0$,  
then the space-time region explored will
be the cone consisting of the points $(x,t)$ such that $t_0\le t \le t_0+L$,
and $|x-x_0| \le v_e t$. (This neglects for the moment the hard 
question of interaction between civilizations.) A point $(x,t)$ will
be explored by some civilization if any civilization starts in the
space-time region 
$$ C_P(x,t)= \{ (y,t-s): 0\le s \le L, |x-y| \le v_e s \}. $$
The space volume explored will initially grow cubically with $L$, but
with a transition to quadratic growth at the time $t_w= h_G/v_e$
taken to cross the thickness of the galactic disc. We have $t_w=10^5$,
and it turns out that it is the case $L \ge t_w$ which is of interest.
The (space-time) volume of $C_P(x,t)$ is of the order of
$$ W_C = \frac{\pi}{3} h_G v_e^2 L^3; $$
the exact value will depend on its location within the galaxy.
The volume of the galaxy is $V_G = \pi R_G^2 h_G$, and so the 
mean number of civilizations arising in the region $C_P(x,t)$ is
$$ M= \frac{\lam W_C}{V_G} = \lam \frac{ h_G v_e^2 L^3}{ 3 h_G R_G^2}
=  \frac{L^3}{T} \frac{v_e^2}{3 R_G^2}.  $$
Taking $3 R_G^2 = 7.5 * 10^9 \simeq 10^{10}$ $ly^3$, we have
$M\ge 1$ when
$$ 3 \log L \ge 14 + \log T. $$ 
(Note that $\log T\ge 1$ then gives $L \ge t_w$.)
If $M \gg 1$ then a typical space-time point in the galaxy will lie in
the exploration cone of many civilizations, and so these
cones will cover most of the galaxy, while if $M \ll 1$ then there will be 
substantial vacant unexplored regions. 

\sm ($R_4$) (`Multiple zones')
In the region $3\log L \ge 14+ \log T$, $\log T \le 7$ we therefore 
expect that the galaxy will covered by the zones of control of more
than civilization. How these civilizations might interact is considered
briefly below.

If $3\log L \le 14+ \log T$ then civilizations are too rare and 
short-lived for their SRPs to cover the galaxy, but we can still ask 
about their  radio signals.
Let us begin by considering the conditions for 2-way communication by radio
with an ETI. 
The same analysis as with the SRPs applies in this case, 
but with $v_e$ replaced by the speed of light $v_c =1$. 
Assume for simplicity that the time between
a civilization starting to send out radio transmissions and sending out 
SRPs is small, and that radio transmissions continue for the lifetime 
of a civilization. Then the mean number
of civilization still extant whose broadcasts can be accessed at a 
point $(t,x)$ will be
$$ M' =  \frac{L^3}{T} \frac{v_c^2}{3 R_G^2} = \frac{M}{v_e^2}. $$
Thus $M' \ge 1$ if $ 3 \log L \ge 10 + \log T. $
(If $\log T\ge 1$ then this condition gives $L \ge 10^{11/3}> 1000$, 
so the case when we
need to consider zones with radius less than $h_G$ does not arise.)

\sm ($R_5$) (`2 way SETI') 
If $10 + \log T \le 3 \log L \le 14+ \log T$ and $\log T \ge 1$
then a typical point will be able to receive radio signals from a civilization
which is still extant, but will not be visited by SRPs. 
There is therefore the possibility of 2-way communication by radio between
two civilizations, possibly continuing until one becomes extinct. 
This is the situation envisaged in much of the early SETI literature.

\sm ($R_6$) (`1-way SETI'). If $3 \log L \le 10+\log T$ and  $\log T \ge 1$
then a typical point can only receive signals from extinct civilizations. 
A point $(t,x)$ will be able to receive signals from a civilization if
that civilization arose in the region
$$ C_S(t,x)= \{ (y,s):   t-|x-y|-L \le s t -|x-y| \}. $$
This has space time volume $L V_G$, and so the mean number of 
such civilizations is $\lam L V_G/V_G = L/T$.
Thus $L\ge T$ is (not surprisingly) also the condition for there to be 
some civilization in the galaxy within our light cone.

The space of galactic ecologies is therefore divided into six
regions. For regions $R_1$ and $R_2$ 
there is little more to be said, but some
other cases deserve further attention.

In region $R_4$ a typical point in the galaxy could be explored by 
SRPs from many civilizations, and it is necessary to consider how
such civilizations might interact. 
One can identify three broad possibilities: \\
(i) No interaction, and mutual interpenetration between explored regions
of different civilizations, \\
(ii) Civilizations establish boundaries between their different
`zones of control', \\
(iii) Civilizations (or their SRPs) engage in warfare. 

In case (i) we would expect to see many probes within our solar system, and
our failure to do so tends towards excluding this possibility. 

For case (ii), consider the arrival of a SRP from Civilization X in a star
system already containing infrastructure 
built by Civilization Y. The probe
would need to decelerate from 0.01c, and this would require the expenditure
of large amounts of energy over a significant period,  making the arrival
detectable by Y.
On arrival the SRP would have limited fuel and resources, and
could be quarantined or neutralised by Y. A (lengthy) period of negotiation
might then lead to agreed boundaries between X and Y.

If negotiation failed then war might ensue, which is case (iii). 
In the case of all out war, constraints on the number of SRPs built
would be dropped, and all available material would be used. 
If it is the case that the material in stars and gas giants is too tightly 
bound gravitationally to be used to make SRPs,
then the effects of such a war on other star systems might not 
be detectable to us at present.  
However, two pieces of evidence support the conclusion 
that such a war has never occurred in our galaxy.
The first is that the solar system has not been mined in this way.
Second, if SRPs can only utilize smaller planets then 
the total mass usable for SRPs in a typical stellar system
would be around $10^{22}- 10^{23}$ kg. However, a protostellar nebula
contains a mass of around $10^{30}$ kg, which is 
not be tightly bound gravitationally.
Such nebulae would be major military prizes, and their continued 
existence in our galaxy, as well as that of recently formed stars, 
suggests that our galaxy has seen neither an all out war, nor an arms race.
(This applies also to other galaxies.)

\section{Conclusion}

Under our hypothesis that the technologies (T1) and (T2) can be attained,
consideration of the points above, and Figure 1, leads to three broad categories
of answer to Fermi's question: 

\sm (F1) They have not visited us because they do not exist. 
(Regions $R_1$ and $R_2$.) \\
(F2) The `zoo hypothesis': their probes are watching us now 
(Regions $R_3$ and $R_4$.) \\
(F3) They have not visited us because civilizations are all too short lived 
(Regions $R_5$ and $R_6$).

Of these, possibility (F3) relies all \textit{all} civilizations being 
short lived, while the zoo hypothesis appears to be deeply unpopular 
(partly I suspect because it compromises human dignity.)
The analysis above reduces the force of some of the objections 
that have been made to the zoo hypothesis, since 
in both cases $R_3$ and $R_4$(ii) 
we would lie in the zone of control of just one ETI.

If we exclude (F2) and (F3), then we are left with (F1), to which there are
no objections except that it is uninteresting. 
It is worth noting that while astronomers have frequently given rather
large values to $f_{ci}$ -- typically in the range $0.01$--$0.1$, 
many evolutionary biologists have been much more pessimistic. 
Even if one is not convinced by all the arguments in 
Ward and Brownlee 2000, it seems very possible that the development
of intelligent life requires evolution to pass through several gateways, 
and hence that $f_{ci}$ is very small.

 \section{Acknowledgements}%\nonumber
 This research was partially supported by NSERC (Canada) 
 and Trinity College Cambridge (UK).

\bibliographystyle{model2-names}
%\bibliography{bibliography}

\end{document}